# Artificial Intelligence (AI) Maturity in Small and Medium-Sized Enterprises: A Framework of Internalized and Ecosystem-Embedded Capabilities

Sukanlaya Sawang[1] and Virach Sornlertlamvanich[2]

## Abstract

Artificial intelligence (AI) maturity models have proliferated, yet prevailing frameworks remain largely enterprise-centric, linear, and weakly aligned with the organizational realities of small and medium-sized enterprises (SMEs). This study develops a conceptual AI maturity framework explicitly grounded in SME contexts. Drawing on organizational capability theory, maturity model research, and the SME digital transformation literature, the framework reconceptualizes AI maturity as a multidimensional, non-linear, and ecosystem-embedded capability. It comprises eight interrelated capability dimensions, five maturity levels, and four archetypal development pathways, capturing heterogeneity in SME AI adoption trajectories. By foregrounding resource constraints, informal governance, owner–manager dominance, and external ecosystem dependence, the framework extends existing AI maturity theory and responds to calls for context-sensitive conceptualizations of AI capability development. The study provides a foundation for future empirical validation and comparative research on AI maturity in SMEs, to measure their competitiveness, potential in self-development, and driving force in SMEs context.

[1] Email: s.sawang@napier.ac.uk, Centre for Business Innovation and Sustainable Solutions, Edinburgh Napier Unievrsity, UK
[2] Email: virach@musashino-u.ac.jp, Faculty of Data Science, Musashino University, Japan

## 1. Introduction

Artificial intelligence (AI) is increasingly recognized as a critical driver of organizational competitiveness, innovation, and productivity. Advances in machine learning, natural language processing, and cloud-based AI services have lowered barriers to adoption, enabling organizations of varying sizes to experiment with AI-enabled solutions (Yang, 2022). Consequently, AI has become increasingly relevant for small and medium-sized enterprises (SMEs), which constitute the backbone of most economies worldwide (Sawang et al., 2024; Wei & Pardo, 2022). Despite this growing accessibility, organizations vary substantially in their ability to adopt, scale, and derive value from AI. While some firms successfully embed AI into core processes and decision-making, many struggle to move beyond isolated pilots or experimental applications (Bedué & Fritzsche, 2022). This uneven diffusion has shifted scholarly attention from whether AI is adopted to how organizations develop the capabilities required to use AI effectively and sustainably. In response, the concept of **AI maturity** has emerged as a lens for examining organizational readiness, capability development, and AI-enabled value creation (Mikalef et al., 2019; Fosso Wamba et al., 2024).

Existing AI maturity models typically conceptualize AI adoption as a staged and internally driven process shaped by dimensions such as data readiness, skills, infrastructure, and governance (Ransbotham et al., 2020). However, these models remain largely enterprise-centric, implicitly assuming formalized governance, specialized AI expertise, and sustained access to resources. Such assumptions are misaligned with SME contexts, where AI adoption is often experimental, uneven, and strongly shaped by resource constraints and reliance on external actors. Research on SME digital transformation highlights that SMEs are not scaled-down versions of large organizations. Instead, they are characterized by informal governance structures, strong owner–manager influence, limited internal specialization, and pronounced dependence on external ecosystems, including technology vendors, consultants, and cloud platforms (Kraus et al., 2021; Torrès & Julien, 2005). These characteristics are theoretically revealing, as they fundamentally alter how AI capabilities are developed, governed, and sustained. Yet, prevailing AI maturity frameworks have largely abstracted away from these conditions, reinforcing a gap between maturity theory and SME realities.

Against this backdrop, this study develops a conceptual AI maturity framework tailored to SME contexts, drawing on a systematic synthesis of academic literature on AI, digital maturity, and SME transformation, complemented by practitioner-oriented AI maturity models. Rather than proposing another linear maturity scale, the framework offers a multidimensional and context-sensitive perspective on AI capability development in SMEs. The primary theoretical contribution of this study is the reconceptualization of AI maturity as a configuration of partially internalized and ecosystem-embedded organizational capabilities, rather than as a linear trajectory of internally accumulated technological sophistication. By shifting the unit of analysis from stages of technological advancement to configurations of organizational and ecosystem capabilities, the framework challenges firm-centric assumptions in AI maturity theory and explains heterogeneous AI development pathways in resource-constrained contexts. This reconceptualization is further elaborated through AI maturity dimensions, maturity levels, and archetypal development pathways that capture non-linear and alternative trajectories of AI capability development in SMEs.

## 2. Theoretical Background

The development of an AI maturity framework for SMEs requires a robust theoretical grounding that moves beyond tool-centric or consultancy-driven perspectives. This study draws on organizational capability theory, particularly the Resource-Based View (RBV) (Barney, 1991; Barney et al., 2011) and Dynamic Capabilities Theory (Teece, 2018; Teece et al., 1997), and integrates these perspectives with maturity models as a theoretical lens for understanding capability development over time. Together, these foundations support a conceptualization of AI maturity as an evolving organizational capability rather than a static measure of technological adoption.

### 2.1 AI as an Organizational Capability

Early discussions of Artificial Intelligence (AI) adoption often conceptualized AI as a discrete technological artifact—such as machine learning algorithms, automation tools, or decision-support systems. However, growing academic consensus suggests that AI should instead be understood as a bundled organizational capability that emerges from the coordinated deployment of technological, human, and organizational resources (Fosso Wamba et al., 2024; Mikalef et al., 2019).

The RBV provides an important theoretical foundation for this perspective. RBV argues that sustained competitive advantage arises not from technology itself, but from firm-specific resources that are valuable, rare, inimitable, and non-substitutable (Barney, 1991; Barney et al., 2011). From an RBV standpoint, AI technologies alone are unlikely to generate durable advantage, as algorithms, software platforms, and cloud-based AI services are increasingly commoditized and widely accessible. Instead, value from AI arises when these technologies are combined with complementary resources such as proprietary data, domain expertise, managerial cognition, and organizational processes (Mikalef et al., 2019; Raisch & Krakowski, 2021).

While RBV explains *what* resources matter, it is limited in explaining *how* organizations adapt these resources in rapidly changing environments. This limitation is addressed by Dynamic Capabilities Theory, which emphasizes a firm's ability to sense opportunities, seize them, and reconfigure resources in response to environmental change (Teece, 2018; Teece et al., 1997). AI adoption is inherently dynamic: models must be continuously trained, data pipelines refined, governance mechanisms updated, and human–AI interaction recalibrated. As such, AI maturity reflects not a static endpoint but an organization's ongoing capacity to learn, adapt, and scale AI capabilities over time. Empirical research supports this capability-based view of AI. Studies show that organizational performance from AI depends on complementary capabilities such as data governance, analytical skills, leadership support, and cross-functional integration, rather than the sheer number of AI applications deployed (Mikalef et al., 2019; Wamba et al., 2021). This insight is particularly relevant for SMEs, where resource constraints mean that selective, well-aligned capability development is more critical than broad technological adoption.

Accordingly, AI maturity cannot be equated with the number or sophistication of AI tools in use. An organization deploying multiple AI applications without strategic alignment, skilled personnel, or governance mechanisms may exhibit lower AI maturity than a firm using fewer tools but embedding them effectively into core processes. This distinction underpins the conceptualization of AI maturity adopted in this study.

## 2.2 Maturity Models as a Theoretical Lens

Maturity models offer a structured theoretical approach for examining how organizational capabilities develop over time (Becker et al., 2009). Originating in software engineering, most notably through the Capability Maturity Model (CMM), maturity models conceptualize development as a progression from informal, ad hoc practices toward more structured, integrated, and optimized capabilities (Paulk et al., 1991). Since then, maturity models have been widely applied in domains such as information systems, digital transformation, analytics, and Industry 4.0.

Conceptually, maturity models can be understood along two key dimensions. *First*, they may adopt a staged or evolutionary logic. Staged models describe capability development through discrete levels with identifiable characteristics, whereas evolutionary models emphasize gradual, non-linear progression and learning over time. While staged representations are often used for clarity and assessment, contemporary research recognizes that organizations—particularly SMEs—may progress unevenly across dimensions and may temporarily regress or plateau (Sadiq et al., 2021)(Sadiq et al., 2021). *Second*, maturity models serve both diagnostic and prescriptive functions. Diagnostically, they allow organizations to assess their current state relative to a set of capability dimensions. Prescriptively, they provide guidance on feasible improvement paths and capability-building priorities (Becker et al., 2009). This dual role makes maturity models especially suitable for complex, multi-dimensional phenomena such as AI adoption, where organizations require both assessment and direction.

In AI adoption research, maturity models are increasingly used to capture the interaction between technology, organization, and people. Systematic reviews of AI maturity models indicate that most frameworks incorporate dimensions such as strategy, data, skills, governance, and processes, reflecting a socio-technical understanding of AI (Leão & da Silva, 2021). However, many existing models remain practitioner-driven and lack explicit theoretical grounding, leading to critiques that they reflect consultancy logic rather than theory-driven constructs.

The current study addresses that concern by grounding the maturity model explicitly in RBV and Dynamic Capabilities Theory. From this perspective, maturity levels represent configurations of organizational capabilities, rather than simple technological milestones. Maturity progression reflects increasing ability to sense AI opportunities, integrate AI into organizational routines, and reconfigure resources to sustain value creation. Maturity models are therefore particularly appropriate for AI adoption research because they align with the evolutionary nature of capability development emphasized in organizational theory. For SMEs, they provide a pragmatic yet theoretically grounded mechanism for understanding AI adoption as a journey shaped by learning, constraints, and contextual factors rather than as a linear technology rollout.

## 2.3 AI Maturity Models in Existing Literature

The concept of AI maturity has gained increasing attention in both academic and practitioner-oriented research as organizations seek to assess and guide their AI adoption journeys. Although definitions vary, AI maturity models generally conceptualize AI adoption as a multi-dimensional and evolutionary process, reflecting the interplay between technological, organizational, and human factors. Across existing models, five dominant dimensions consistently emerge:

**Strategy** is widely recognized as a foundational dimension of AI maturity. Studies emphasize the importance of aligning AI initiatives with organizational goals, competitive positioning, and value creation logic (Davenport & Ronanki, 2018; Ransbotham et al., 2020). Strategic maturity reflects not only the presence of an AI roadmap but also leadership commitment and clarity regarding AI's role in decision-making and innovation.

**Data** constitutes another central pillar of AI maturity. Academic research consistently identifies data availability, quality, integration, and governance as prerequisites for effective AI deployment (Kiron et al., 2013; Mikalef et al., 2019). AI maturity models emphasize that without robust data foundations, organizations are unable to train, scale, or sustain AI systems, regardless of algorithmic sophistication.

**Technology and infrastructure** form the technical backbone of AI maturity. This dimension encompasses analytics platforms, cloud infrastructure, system integration, and model deployment capabilities (Fosso Wamba et al., 2024). However, research increasingly cautions against equating technological sophistication with maturity, highlighting that technology must be embedded into organizational routines to generate value (Rai et al., 2019).

**People and skills** are repeatedly identified as critical determinants of AI maturity. Studies highlight the importance of AI literacy among managers, availability of analytical talent, and cross-functional collaboration between technical and business roles (Mikalef et al., 2019; Raisch & Krakowski, 2021). AI maturity thus reflects not only technical expertise but also the organization's capacity to interpret, trust, and act upon AI-driven insights.

Finally, **governance and ethics** have emerged as increasingly prominent dimensions in recent AI maturity frameworks. Governance encompasses issues such as accountability, transparency, explainability, risk management, and regulatory compliance (Floridi et al., 2018; Shrestha et al., 2019). As AI systems become more embedded in organizational decision-making, maturity models emphasize the need for responsible AI practices to sustain trust and legitimacy.

Overall, these dimensions demonstrate a strong convergence across AI maturity models, suggesting a shared understanding of AI as a socio-technical organizational capability rather than a purely technological artifact. However, convergence at the dimensional level does not imply adequacy across organizational contexts, particularly for SMEs.

## 2.4 Limitations of Existing AI Maturity Models

Despite their conceptual value, existing AI maturity models exhibit several limitations that constrain their applicability, especially in SME contexts. *First*, a pronounced *enterprise bias* characterizes much of the AI maturity literature. Many frameworks are developed based on large organizations with access to extensive financial resources, specialized teams, and formalized governance structures (Fosso Wamba et al., 2024; Ransbotham et al., 2020). These assumptions often do not hold for SMEs, which operate with lean structures and limited internal specialization. As a result, enterprise-oriented maturity benchmarks may be unrealistic or misleading when applied to smaller firms.

*Second*, many AI maturity models rely on *linear progression assumptions*, depicting maturity as a sequential movement through predefined stages. While such representations offer clarity, they risk oversimplifying the complex and non-linear nature of AI adoption. Empirical studies show that organizations often develop AI capabilities unevenly across dimensions, with experimentation, stagnation, and regression occurring simultaneously (Kane et al., 2017; Pumplun et al., 2019; Ransbotham et al., 2020). This limitation is particularly relevant for SMEs, where opportunistic adoption and project-based experimentation are common.

*Third*, existing models frequently assume *high resource availability*, including advanced data infrastructures, dedicated AI governance bodies, and continuous investment capacity. However, SMEs face chronic constraints in capital, data volume, and specialized talent (OECD, 2021). Consequently, maturity models that equate progress with scale and sophistication risk penalizing SMEs despite effective, context-appropriate AI use.

*Fourth*, there is *limited empirical grounding in SME contexts.* While AI maturity models are increasingly discussed conceptually, relatively few have been empirically validated using SME samples (Pumplun et al., 2019). This lack of empirical grounding raises concerns regarding external validity and further reinforces the need for SME-specific conceptualization. Collectively, these limitations suggest that existing AI maturity models, while conceptually rich, inadequately capture the realities of AI adoption in SMEs.

## 2.5 SME-Specific Digital and AI Adoption Challenges

A defining characteristic of SMEs is the structural scarcity of resources (Sawang & Unsworth, 2011), which fundamentally conditions their digital and AI adoption trajectories. Unlike large enterprises, SMEs typically operate with limited financial slack, small and multifunctional workforces, and constrained access to advanced technological infrastructures (Vial, 2019). These constraints do not merely slow AI adoption; they shape how AI is adopted. Empirical research suggests that SMEs prioritize incremental, low-risk, and problem-specific AI applications rather than organization-wide transformation initiatives (Pumplun et al., 2019; Raisch & Krakowski, 2021; Vial, 2021). Consequently, AI maturity in SMEs is less about scale or technological sophistication and more about selective capability development under constraint—a reality insufficiently captured by many existing maturity models that implicitly assume continuous investment capacity.

Beyond resource constraints, SMEs are distinguished by informal organizational and governance structures. Decision-making authority is often centralized, formal processes are limited, and strategic planning horizons tend to be short and adaptive rather than long-term and analytical (Thakur & Sinha, 2024; Torrès & Julien, 2005). While such informality can enhance responsiveness and flexibility, it complicates the implementation of formalized AI governance mechanisms—such as dedicated oversight bodies, standardized evaluation procedures, and codified ethical guidelines—that feature prominently in enterprise-oriented AI maturity frameworks. This tension suggests that governance maturity in SMEs cannot be equated with formalization but must instead be understood as contextually appropriate control and accountability mechanisms.

A further distinguishing factor is the central role of owner–manager dominance in shaping technology adoption decisions. Upper echelons theory and SME research consistently demonstrate that the beliefs, cognitive frames, and risk perceptions of owner–managers exert a disproportionate influence on digital investment choices (Bresciani et al., 2022; Raymond & Bergeron, 2008). In the context of AI, this implies that maturity is not solely a function of organizational resources or technical readiness, but also of managerial sensemaking, trust in algorithmic outputs, and leadership orientation toward experimentation and learning. Existing AI maturity models, however, largely abstract away from these cognitive and behavioral dynamics.

Finally, SMEs exhibit a high degree of vendor and ecosystem dependence in their AI adoption efforts. Rather than internalizing AI capabilities, SMEs frequently rely on external vendors, consultants, cloud platforms, and public support infrastructures to access AI technologies (Autio et al., 2018). As a result, AI capabilities in SMEs are often distributed across organizational boundaries, embedded within inter-organizational arrangements rather than fully owned or controlled. This externalization challenges the firm-centric assumptions underpinning many AI maturity models and suggests that ecosystem engagement should be treated as a core dimension of AI maturity rather than a peripheral enabling factor.

## 2.6 Limitations of Existing AI Maturity Models and Research Gap

A synthesis of the extant literature reveals a persistent misalignment between dominant AI maturity models and the organizational realities of small and medium-sized enterprises. Although existing frameworks exhibit convergence around core dimensions—such as strategy, data, technology, people, and governance—they are predominantly developed within large-enterprise contexts and frequently embed assumptions of formalized structures, linear progression, and sustained resource availability (Becker et al., 2009; Reichl & Gruenbichler, 2023). These assumptions implicitly position AI maturity as a staged movement toward enterprise-scale sophistication, thereby limiting their applicability to smaller firms.

In contrast, research on digital transformation and technology adoption in SMEs consistently depicts capability development as non-linear, uneven, and path-dependent. SMEs often pursue iterative and experimental adoption trajectories characterized by selective capability development, stalled initiatives, and parallel advancement across some dimensions but not others (Vial, 2019). Such patterns challenge the linear logic underlying many AI maturity models and suggest that maturity may manifest as heterogeneous configurations rather than uniform stages. Moreover, SME research highlights structural characteristics that are largely overlooked in enterprise-oriented maturity frameworks. Informal governance arrangements, centralized decision-making authority, and adaptive strategic planning horizons shape how technologies are evaluated and deployed in smaller firms (Kraus et al., 2020). These features complicate the translation of formal AI governance mechanisms and standardized maturity benchmarks into SME contexts, where control and coordination are often exercised through personal oversight rather than codified processes.

A further limitation of existing AI maturity models lies in their largely firm-centric orientation, which underestimates the role of external ecosystems in capability development. SMEs frequently rely on external vendors, consultants, cloud platforms, and public support infrastructures to access AI technologies, resulting in AI capabilities that are partially externalized and distributed across organizational boundaries (Autio et al., 2018). Recent reviews of AI and digital transformation research emphasize that such ecosystem dependence is particularly pronounced among SMEs, yet remains insufficiently theorized in prevailing AI maturity frameworks (Cenamor et al., 2017).

Overall, these limitations point to a substantive gap in the AI maturity literature. There is a lack of frameworks that conceptualize AI maturity as context-sensitive, non-linear, and ecosystem-embedded, reflecting the distinctive structural, cognitive, and resource conditions of SMEs. Addressing this gap requires moving beyond linear, enterprise-centric models toward a capability-based perspective that accommodates heterogeneous adoption pathways and recognizes that AI maturity in SMEs may be achieved through flexible configurations of internal and external capabilities. Responding to this need forms the foundation for the conceptual AI maturity framework developed in the subsequent sections. Table 1 summarizes the key distinctions between dominant AI maturity models and the proposed framework, targeting on SMEs potential in self-development and competitiveness in business.

**Table 1.** Comparison of Existing AI Maturity Models and the Proposed SME-Focused Framework

| Dimension | Existing AI Maturity Models | Proposed AI Maturity Framework for SMEs |
|---|---|---|
| Primary focus | Scaling AI sophistication and optimization | Building fit-for-purpose AI capabilities |
| Underlying assumptions | Abundant resources, formal structures, internal AI teams | Resource constraints, informality, external support |
| Progression logic | Linear, stage-based advancement | Non-linear, contextual, capability-based |
| View of AI | Technology-driven or process-driven | Bundled organizational capability |
| Governance orientation | Formalized, enterprise-level controls | Scaled, pragmatic, context-sensitive governance |
| Role of external ecosystem | Largely implicit or ignored | Explicitly integrated (vendors, platforms, policy) |
| Target firms | Large enterprises, digitally mature organizations | Small and medium-sized enterprises (SMEs) |
| Intended use | Benchmarking and enterprise transformation | Diagnostic guidance and realistic capability development |

## 3. Conceptualization of AI Maturity in SME Contexts

This section presents the conceptual AI maturity framework developed in this study. The framework is articulated at an abstract level, emphasizing theoretical integration and analytical clarity rather than exhaustive operational detail. Consistent with an organizational capability perspective, AI maturity is conceptualized as the extent to which an SME can purposefully deploy, integrate, and continuously adapt AI-enabled solutions in pursuit of business objectives. Rather than treating maturity as a function of technological sophistication alone, the framework conceptualizes AI maturity as an emergent configuration of organizational and ecosystem capabilities.

The framework is explicitly grounded in SME contexts. Across all dimensions, AI maturity in SMEs is shaped by persistent resource constraints, informal governance structures, strong owner–manager influence, and reliance on external ecosystems. As a result, AI maturity does not reflect scaled-down versions of enterprise practices, but context-specific manifestations of capability development under constraint. Structurally, the framework comprises (1) a set of core AI maturity dimensions representing key organizational capability domains, (2) maturity levels capturing progressive stages of AI capability development, and (3) archetypal development pathways reflecting heterogeneity in how SMEs combine breadth and depth of AI adoption. While maturity levels are presented sequentially for analytical clarity, progression is explicitly non-linear and may vary across dimensions.

### 3.1 Core Dimensions of AI Maturity in SMEs

### 3.1.1 Strategic Orientation and Leadership

This dimension captures the extent to which AI-related strategic intent and leadership commitment guide organizational decision-making and capability development (Porfírio, Carrilho, Felício & Jardim, 2021). From a capability perspective, strategic orientation shapes how AI initiatives are prioritized, resourced, and aligned with business objectives. In SME contexts, this dimension is strongly influenced by owner–manager cognition and typically manifests through emergent, judgment-based strategies rather than formalized AI roadmaps.

### 3.1.2 Human Capital and AI-Related Competencies

Human capital and competencies refer to the organization's capacity to understand, apply, and learn from AI-enabled solutions (Zhang & Chen, 2024). AI maturity depends not only on technical expertise but also on managerial AI literacy and cross-functional collaboration. In SMEs, this dimension is often characterized by generalist skill profiles, learning-by-doing, and selective reliance on external expertise rather than dedicated AI teams.

### 3.1.3 Data Resources and Technological Foundations

This dimension reflects the accessibility, quality, and usability of data, as well as the technological infrastructure that supports AI development and deployment. From a maturity perspective, data and technology function as enabling conditions rather than sources of value in isolation (Kapoor & Teece, 2021). In SMEs, AI maturity is shaped by the ability to mobilize existing data assets pragmatically, rather than by the presence of enterprise-scale data architectures.

### 3.1.4 AI Application Scope and Business Embedding

This dimension captures the extent to which AI applications are embedded in organizational activities and generate business-relevant outcomes. AI maturity is reflected in the alignment between AI use cases and strategic or operational priorities (Sonntag et al., 2024). SMEs typically focus on a limited number of high-impact applications, emphasizing relevance and effectiveness over portfolio breadth.

### 3.1.5 Process Integration and Operational Alignment

Process integration refers to the degree to which AI-enabled solutions are embedded within routine workflows and operational processes. AI maturity emerges when AI outputs are systematically incorporated into decision-making and execution (Davenport & Ronanki, 2018). In SME contexts, even limited AI integration can yield disproportionate operational benefits, provided solutions are aligned with existing routines.

### 3.1.6 Technical Sophistication and Solution Appropriateness

This dimension reflects the complexity and customization of AI solutions relative to organizational needs and capabilities. AI maturity does not imply maximal technical sophistication; rather, it reflects the deployment of solutions whose complexity is appropriate to the problem context and organizational capacity (Cohen & Levinthal, 1990; Raisch &

Krakowski, 2021)). For SMEs, right-sized and adaptable solutions are often more sustainable than highly advanced or bespoke systems.

### 3.1.7 Performance Evaluation and Learning Mechanisms

This dimension captures the extent to which SMEs monitor AI performance and engage in continuous learning and improvement. AI maturity depends on the ability to evaluate outcomes, generate feedback, and refine solutions over time (Sonntag et al., 2024). In SMEs, learning mechanisms tend to be lightweight and outcome-oriented, supporting adaptation despite limited analytical resources.

### 3.1.8 Risk Governance and Responsible AI Practices

Risk governance refers to how organizations manage ethical, legal, and operational risks associated with AI use. AI maturity entails awareness of accountability, transparency, and regulatory considerations, even in the absence of formal governance structures (Krijger, et al., 2023). In SMEs, responsible AI practices are typically principle-based and pragmatic rather than codified through formal oversight bodies.

### 3.2 Interdependencies Across AI Maturity Dimensions in SMEs

Although the proposed dimensions are analytically distinct, AI maturity in SMEs emerges through their interaction rather than through independent development. Strategic orientation shapes how data, skills, and technologies are prioritized, while human competencies condition the organization's ability to integrate AI into operational processes. Similarly, data readiness and technological foundations enable, but do not determine the effective AI use unless embedded within organizational routines and supported by appropriate governance mechanisms. Importantly, reliance on external ecosystems cuts across all dimensions, influencing strategic choices, skill development, solution complexity, and risk management practices. These interdependencies reinforce the conceptualization of AI maturity as a configuration of mutually reinforcing organizational and ecosystem capabilities rather than a collection of discrete attributes (Jacobides, Cennamo & Gawer, 2018).

### 4. Reconceptualizing Progression: Levels and Archetypes

AI maturity levels provide a structured representation of progressive capability development. The framework distinguishes five levels: *Discovery, Experimentation, Implementation, Deployment, and Optimization*. These levels represent increasing degrees of organizational learning, integration, and refinement. While presented sequentially, progression across levels is not assumed to be uniform or irreversible. SMEs may advance unevenly across dimensions, stabilize at intermediate levels, or selectively optimize AI capabilities in specific domains without pursuing organization-wide transformation.

To capture heterogeneity in AI development trajectories, the framework introduces AI maturity archetypes that reflect distinct configurations of organizational and ecosystem capabilities in SMEs. While maturity levels provide an analytically useful representation of progressive capability development, they are limited in explaining why SMEs at similar maturity levels often exhibit fundamentally different AI outcomes. Archetypes address this limitation by shifting the analytical focus from *where* an organization is positioned along a maturity continuum to *how* AI-related capabilities are configured, coordinated, and governed.

Building on configuration theory (Fiss, 2007), archetypes represent internally consistent patterns through which SMEs combine strategic orientation, internal capabilities, and external ecosystem resources under conditions of constraint. In this sense, archetypes perform explanatory work that maturity levels alone cannot: they account for equifinality, whereby multiple capability configurations can lead to comparable levels of AI-enabled value creation despite differences in internalization, integration, and governance.

The framework identifies four archetypal AI maturity pathways commonly observed in SME contexts. *Emerging Explorers* are characterized by fragmented and exploratory AI engagement, relying on limited internal capabilities and ad hoc experimentation shaped primarily by owner–manager interest. *Broad Implementers* deploy AI-enabled solutions across multiple organizational activities through standardized, vendor-provided, or platform-based technologies; although functional AI use is achieved, internal capability development and cross-initiative coordination remain limited. *Focused Specialists* concentrate AI capability development within strategically salient domains, combining selective internal learning with targeted ecosystem support to achieve deep integration in specific processes. *Advanced Leaders* exhibit internally coherent and externally coordinated AI capability configurations, enabling sustained integration of AI into strategic decision-making and operational routines.

Crucially, archetypes also illuminate strategic choice versus structural constraint in SME AI development. Some archetypal configurations reflect deliberate strategic focus (e.g., Focused Specialists), whereas others emerge primarily from resource limitations or ecosystem dependence (e.g., Broad Implementers). Moreover, archetypes enable theorization of stability and transition: SMEs may remain stably embedded within a given archetype for extended periods, or transition between archetypes as internal capabilities accumulate, governance mechanisms evolve, or ecosystem relationships are reconfigured. Such transitions are not necessarily upward along maturity levels but may involve lateral reconfiguration of capabilities. By incorporating archetypes, the framework moves beyond linear maturity assumptions and positions AI maturity as a configuration-based and path-dependent phenomenon. Archetypes thus constitute a core theoretical component of the framework, enabling explanation of heterogeneity, non-linearity, and divergent AI development pathways in SMEs. This perspective provides a foundation for future empirical research examining how different archetypal configurations influence AI performance outcomes, organizational learning, and long-term adaptability.

**Table 2**. Integrated View of the AI Maturity Framework for SMEs

| Framework Element | Conceptual Role | Description |
|---|---|---|
| Core AI Maturity Dimensions | Capability domains | Eight interrelated organizational and ecosystem capability domains that jointly shape AI maturity in SMEs: strategic orientation and leadership; human capital and AI-related competencies; data resources and technological foundations; AI application scope and business embedding; process integration and operational alignment; solution appropriateness; performance evaluation and learning mechanisms; and risk governance and responsible AI practices. |
| Interdependencies Across Dimensions | Configuration logic | AI maturity emerges through complementarities and interactions among dimensions rather than independent advancement, reflecting a configuration of mutually reinforcing organizational and ecosystem capabilities. |
| AI Maturity Levels | Developmental progression | Five analytically distinct levels—Discovery, Experimentation, Implementation, Deployment, and Optimization—represent progressive stages of AI capability development. Levels provide structural orientation but do not imply uniform, linear, or irreversible progression across dimensions. |
| AI Maturity Archetypes | Development pathways | Archetypes capture recurring patterns in how SMEs configure and coordinate AI capabilities under resource constraints and ecosystem dependence. Archetypes reflect equifinality by illustrating multiple viable pathways to AI-enabled value creation. |
| Emerging Explorers | Archetypal configuration | Characterized by exploratory and fragmented AI engagement, limited internal capabilities, and ad hoc experimentation driven largely by managerial interest rather than formal strategy. |
| Broad Implementers | Archetypal configuration | Characterized by the deployment of AI-enabled solutions across multiple activities through standardized or vendor-provided technologies, with limited internal capability development and weak cross-initiative coordination. |
| Focused Specialists | Archetypal configuration | Characterized by selective internal capability development and concentrated AI integration within strategically salient domains, supported by targeted ecosystem partnerships. |
| Advanced Leaders | Archetypal configuration | Characterized by internally coherent and externally coordinated AI capability configurations, enabling sustained integration of AI into strategic decision-making and operational routines. |

## 5. Discussion

This study advances AI maturity framework by reconceptualizing maturity as a configuration of partially internalized and ecosystem-embedded organizational capabilities. Rather than viewing AI maturity as a linear accumulation of internally owned technological assets (Mullally, 2014; Wade & Hulland, 2004)), the proposed framework emphasizes how organizations—particularly SMEs—combine internal resources with externally sourced capabilities to achieve AI-enabled value creation. This reconceptualization has important implications for how AI maturity is theorized, measured, and compared across organizational contexts. The following discussion elaborates these implications by showing how capability orientation, non-linearity, ecosystem dependence, and SME specificity emerge as logical consequences of this theoretical shift.

### 5.1 AI Maturity as a Configuration of Organizational Capabilities

This subsection elaborates how the proposed framework operationalizes AI maturity as a *capability configuration,* reinforcing the argument that maturity reflects alignment across organizational and ecosystem resources rather than technological depth alone. Prior research emphasizes that digital and AI technologies generate value only when combined with complementary organizational resources, such as managerial cognition, human skills, data governance, and operational processes (Mikalef et al., 2019; Rai et al., 2019). By structuring AI maturity around multiple interrelated capability dimensions, the framework aligns with the resource-based view and dynamic capability theory, which conceptualize competitive advantage as arising from the firm's ability to integrate, reconfigure, and deploy resources over time (Teece et al., 1997; Teece, 2018). This perspective is particularly salient for SMEs, where AI adoption rarely entails large-scale infrastructure investments but instead depends on the purposeful orchestration of limited internal and external resources. Consequently, the framework shifts AI maturity research away from technological benchmarks toward the organizational conditions that enable sustained AI-enabled value creation.

## 5.2 Non-Linearity as a Consequence of Configuration-Based Maturity

The observed non-linearity in AI maturity trajectories follows directly from a configuration-based view of capability development, where different combinations of internal and external resources can yield functionally equivalent outcomes. Traditional maturity models assume cumulative and irreversible progression through predefined stages (Becker et al., 2009). In contrast, research on digital transformation demonstrates that organizations—particularly SMEs—often experience uneven, experimental, and reversible development paths (Kane et al., 2017; Vial, 2021). Within the proposed framework, non-linearity is not treated as an anomaly but as an inherent property of configuration-based maturity, reflecting selective capability development, shifting strategic priorities, and changing ecosystem relationships.

## 5.3 Ecosystem Embeddedness as a Structural Condition of AI Maturity in SMEs

From a configuration perspective, ecosystem dependence emerges as a structural condition rather than a contextual add-on in SME AI maturity. Existing AI maturity models largely assume that capabilities are internally developed and governed. However, SMEs frequently rely on external vendors, cloud platforms, consultants, and public support infrastructures to access AI technologies and expertise (Autio et al., 2018; OECD, 2021). The proposed framework incorporates this externalization directly into the conceptualization of AI maturity, recognizing that internal and external capabilities jointly shape AI outcomes. Importantly, the proposed AI maturity archetypes should therefore not be interpreted as alternative maturity models, but as analytically useful representations of distinct capability configurations that emerge under varying resource and ecosystem conditions.

## 5.4 Contributions to SME Digital Transformation Research

The framework also contributes to SME digital transformation research by demonstrating how organizational size, governance, and leadership structures shape AI capability development through configuration effects. Prior studies highlight that SMEs are characterized by resource constraints, informal governance, and strong owner–manager influence, which fundamentally condition technology adoption decisions (Kraus et al., 2021; Raymond & Bergeron, 2008). By embedding these characteristics into the conceptualization of AI maturity, the framework moves beyond treating SMEs as scaled-down enterprises and instead positions them as a theory-revealing context for understanding configuration-based digital transformation. AI maturity in SMEs is thus best understood in terms of contextual fit, strategic focus, and learning capacity rather than convergence toward enterprise-scale ideals.

## 5.5 Practical Implications

From a practical perspective, the framework offers SME managers a diagnostic lens for assessing AI maturity without benchmarking themselves against unrealistic enterprise standards. By highlighting multiple maturity dimensions and archetypal pathways, the framework enables managers to identify strategically relevant capability gaps and pursue AI adoption in an incremental and context-appropriate manner. For policymakers and ecosystem actors, the findings underscore the importance of shared infrastructures, advisory services, and ecosystem orchestration in supporting SME AI maturity. Rather than focusing solely on firm-level technology adoption, policy interventions may be more effective when they strengthen the ecosystems through which SMEs access AI capabilities. Finally, consultants and technology vendors can use the framework to tailor AI solutions to SME maturity profiles, avoiding over-engineered implementations that exceed organizational capacity and undermine value realization.

## 6. Conclusion

The rapid diffusion of Artificial Intelligence has intensified the need for robust frameworks that help organizations understand and develop AI-related capabilities. While AI maturity models have proliferated in recent years, most remain enterprise-centric and insufficiently attuned to the realities of small and medium-sized enterprises. This study addressed this gap by developing a conceptual AI maturity framework specifically tailored to SMEs. Grounded in organizational capability theory and maturity model research, the framework reconceptualizes AI maturity as a multi-dimensional, evolutionary capability encompassing strategy and leadership, people and skills, data and infrastructure, applications and use cases, processes and operations, solution complexity, measurement and improvement, and risk and compliance. By integrating maturity levels and archetypes, the framework captures both progression and heterogeneity in SME AI adoption pathways.

## 7. References

Autio, E., Nambisan, S., Thomas, L. D., & Wright, M. (2018). Digital affordances, spatial affordances, and the genesis of entrepreneurial ecosystems. *Strategic Entrepreneurship Journal, 12*(1), 72-95.

Barney, J. (1991). Firm resources and sustained competitive advantage. *Journal of management, 17*(1), 99-120. https://doi.org/10.1177/014920639101700108

Barney, J. B., Ketchen Jr, D. J., & Wright, M. (2011). The future of resource-based theory: revitalization or decline? *Journal of management, 37*(5), 1299-1315. https://doi.org/10.1177/0149206310391805

Becker, J., Knackstedt, R., & Pöppelbuß, J. (2009). Developing maturity models for IT management: A procedure model and its application. *Business & information systems engineering*, *1*(3), 213-222. https://doi.org/10.1007/s12599-009-0044-5

Bedué, P., & Fritzsche, A. (2022). Can we trust AI? An empirical investigation of trust requirements and guide to successful AI adoption. *Journal of Enterprise Information Management*, *35*(2), 530-549. https://doi.org/10.1108/JEIM-06-2020-0233

Bresciani, S., Ferraris, A., Santoro, G., & Kotabe, M. (2022). Opening up the black box on digitalization and agility: Key drivers and main outcomes. In (Vol. 178, pp. 121567): Elsevier.

Cenamor, J., Sjödin, D. R., & Parida, V. (2017). Adopting a platform approach in servitization: Leveraging the value of digitalization. *International Journal of Production Economics*, *192*, 54-65. https://doi.org/10.1016/j.ijpe.2016.12.033

Cohen, W. M., & Levinthal, D. A. (1990). Absorptive capacity: A new perspective on learning and innovation. *Administrative science quarterly*, *35*(1), 128-152. https://www.jstor.org/stable/pdf/2393553

Davenport, T. H., & Ronanki, R. (2018). Artificial intelligence for the real world. *Harvard business review*, *96*(1), 108-116. https://hbr.org/2018/01/artificial-intelligence-for-the-real-world

Floridi, L., Cowls, J., Beltrametti, M., Chatila, R., Chazerand, P., Dignum, V., Luetge, C., Madelin, R., Pagallo, U., & Rossi, F. (2018). AI4People—An ethical framework for a good AI society: Opportunities, risks, principles, and recommendations. *Minds and machines*, *28*(4), 689-707. https://doi.org/10.1007/s11023-018-9482-5

Fosso Wamba, S., Queiroz, M. M., Pappas, I. O., & Sullivan, Y. (2024). Artificial intelligence capability and firm performance: a sustainable development perspective by the mediating role of data-driven culture. *Information Systems Frontiers*, *26*(6), 2189-2203. https://doi.org/10.1007/s10796-023-10460-z

Kane, G. C., Palmer, D., Nguyen-Phillips, A., Kiron, D., & Buckley, N. (2017). Achieving digital maturity. *MIT Sloan Management Review*, *59*(1). https://sloanreview.mit.edu/projects/achieving-digital-maturity/

Kapoor, R., & Teece, D. J. (2021). Three faces of technology's value creation: Emerging, enabling, embedding. *Strategy Science*, *6*(1), 1-4. https://doi.org/10.1287/stsc.2021.0124

Kiron, D., Prentice, P. K., & Ferguson, R. B. (2013). Raising the bar with analytics. *MIT Sloan Management Review*. https://sloanreview.mit.edu/article/raising-the-bar-with-analytics/

Kraus, S., Jones, P., Kailer, N., Weinmann, A., Chaparro-Banegas, N., & Roig-Tierno, N. (2021). Digital Transformation: An Overview of the Current State of the Art of Research. *Sage Open*, *11*(3), 21582440211047576. https://doi.org/10.1177/21582440211047576

Krijger, J., Thuis, T., de Ruiter, M. *et al.* The AI ethics maturity model: a holistic approach to advancing ethical data science in organizations. *AI Ethics* **3**, 355–367 (2023). https://doi.org/10.1007/s43681-022-00228-7

Leão, P., & da Silva, M. M. (2021). Impacts of digital transformation on firms' competitive advantages: A systematic literature review. *Strategic Change*, *30*(5), 421-441. https://doi.org/10.1002/jsc.2459

Mikalef, P., Boura, M., Lekakos, G., & Krogstie, J. (2019). Big data analytics capabilities and innovation: the mediating role of dynamic capabilities and moderating effect of the

environment. *British journal of management*, *30*(2), 272-298. https://doi.org/10.1111/1467-8551.12343

OECD. (2021). *The Digital Transformation of SMEs, OECD Studies on SMEs and Entrepreneurship*.

Paulk, M. C., Curtis, B., & Chrissis, M. B. (1991). *Capability maturity model for software*.

Porfirio, J. A., Carrilho, T., Felício, J. A., & Jardim, J. (2021). Leadership characteristics and digital transformation. *Journal of Business Research*, *124*, 610-619. https://doi.org/10.1016/j.jbusres.2020.10.058

Pumplun, L., Tauchert, C., & Heidt, M. (2019). A new organizational chassis for artificial intelligence-exploring organizational readiness factors. The 27th European Conference on Information Systems (ECIS), Stockholm & Uppsala, Sweden.

Rai, A., Constantinides, P., & Sarker, S. (2019). Editor's comments: Next-generation digital platforms: Toward human–AI hybrids. In (Vol. 43, pp. iii-ix): Management Information Systems Research Center, University of Minnesota.

Raisch, S., & Krakowski, S. (2021). Artificial intelligence and management: The automation–augmentation paradox. *Academy of management review*, *46*(1), 192-210. https://doi.org/10.5465/amr.2018.0072

Ransbotham, S., Khodabandeh, S., Kiron, D., Candelon, F., Chu, M., & LaFountain, B. (2020). *Expanding AI's impact with organizational learning*. https://web-assets.bcg.com/f1/79/cf4f7dce459686cfee20edf3117c/mit-bcg-expanding-ai-impact-with-organizational-learning-oct-2020.pdf

Raymond, L., & Bergeron, F. (2008). Enabling the business strategy of SMEs through e-business capabilities: a strategic alignment perspective. *Industrial Management & Data Systems*, *108*(5), 577-595. https://doi.org/10.1108/02635570810876723

Reichl, G., & Gruenbichler, R. (2023). Maturity Models for the use of Artificial Intelligence in Enterprises: A Literature Review. The 19th International Scientific Conference on Industrial Systems. ,

Sadiq, R. B., Safie, N., Abd Rahman, A. H., & Goudarzi, S. (2021). Artificial intelligence maturity model: a systematic literature review. *PeerJ Computer Science*, *7*, e661.

Sawang, S., Ng, P. Y., Kivits, R. A., Dsilva, J., & Locke, J. (2024). Examining the influence of customers, suppliers, and regulators on environmental practices of SMEs: Evidence from the United Arab Emirates. *Business Strategy and the Environment*, *33*(7), 6533-6546. https://doi.org/10.1002/bse.3831

Sawang, S., & Unsworth, K. L. (2011). A model of organizational innovation implementation effectiveness in small to medium firms. *International Journal of Innovation Management*, *15*(05), 989-1011. https://doi.org/10.1142/S1363919611003398

Shrestha, Y. R., Ben-Menahem, S. M., & Von Krogh, G. (2019). Organizational decision-making structures in the age of artificial intelligence. *California management review*, *61*(4), 66-83. https://doi.org/10.1177/0008125619862257

Sonntag, M., Mehmann, S., Mehmann, J., & Teuteberg, F. (2024). Development and evaluation of a maturity model for AI deployment capability of manufacturing companies. *Information Systems Management*, *42*(1), 37-67. https://doi.org/10.1080/10580530.2024.2319041

Teece, D. J. (2018). Business models and dynamic capabilities. *Long range planning*, *51*(1), 40-49. https://doi.org/10.1016/j.lrp.2017.06.007

Teece, D. J., Pisano, G., & Shuen, A. (1997). Dynamic capabilities and strategic management. *Strategic management journal*, *18*(7), 509-533. https://doi.org/10.1002/(SICI)1097-0266(199708)18:7<509::AID-SMJ882>3.0.CO;2-Z

Thakur, V., & Sinha, S. (2024). Family governance structures in family businesses: A systematic literature review and future research agenda. *Journal of Small Business Management*, *62*(6), 3016-3052. https://doi.org/10.1080/00472778.2023.2284930

Torrès, O., & Julien, P.-A. (2005). Specificity and denaturing of small business. *International small business journal*, *23*(4), 355-377. https://doi.org/10.1177/02662426050540

Vial, G. (2021). Understanding digital transformation: A review and a research agenda. In Andreas Hinterhuber, Tiziano Vescovi, & F. Checchinato (Eds.), *Managing digital transformation* (pp. 13-66). Routledge.

Wei, R., & Pardo, C. (2022). Artificial intelligence and SMEs: How can B2B SMEs leverage AI platforms to integrate AI technologies? *Industrial Marketing Management*, *107*, 466-483. https://doi.org/10.1016/j.indmarman.2022.10.008

Yang, C.-H. (2022). How artificial intelligence technology affects productivity and employment: Firm-level evidence from Taiwan. *Research Policy*, *51*(6), 104536. https://doi.org/10.1016/j.respol.2022.104536

Zhang, J., & Chen, Z. (2024). Exploring human resource management digital transformation in the digital age. *Journal of the knowledge economy*, *15*(1), 1482-1498. https://doi.org/10.1007/s13132-023-01214-y